\documentstyle[epsfig, 12pt]{article}          
\def\word#1{\,\,\mbox{#1}\,\,}
\def\reff#1{(\ref{#1})}           
            \def\beq{\begin{equation}}
\def\eeq#1{\label{#1}\end{equation}}            
    \def\dfrac#1#2{{\displaystyle\frac{#1}{#2}}}
 
\begin{document}

\footnotetext[1]{Groupe de Physique des 
Particules,D\'epartement de  Physique, Universit\'e  de  Montr\'eal, C.P. 6128, succ. centreville, 
Montr\'eal, Qu\'ebec, Canada, H3C 3J7.}

\footnotetext[2]{present address: Department of Physics, McGill
University, 3600 University St., Montr\'eal, Qu\'ebec, 
H3A-2T8 Canada. E-mail: edery@hep.physics.mcgill.ca}

\title{Causal Structure of Vacuum Solutions to Conformal(Weyl) Gravity}

\author{A. Edery${}^{1,2}$ and M. B. Paranjape${}^{1}$}

\date{ }

\maketitle

\begin{abstract} 
Using Penrose diagrams the causal structure of the static spherically
symmetric vacuum solution to conformal (Weyl) gravity is investigated. A striking aspect of the
solution is an unexpected physical singularity at $r=0$ caused by a linear
term in the metric. We explain how to calculate the deflection of light in coordinates where the metric
is manifestly conformal to flat i.e. in coordinates where light moves in straight lines.
\end{abstract}                    

\centerline{KEY WORDS: Conformal gravity; Penrose diagrams}  
\section{INTRODUCTION}
 
Conformal gravity is a higher derivative metric theory of gravity whose action is given by the Weyl tensor
squared,
\beq
I=\alpha\int \sqrt{-g} \, C_{\lambda\mu\sigma\tau}C^{\lambda\mu\sigma\tau} d^4 x
\eeq{action} 
where $\alpha$ is a dimensionless constant. It is the simplest action that can be constructed which is conformally
invariant i.e. invariant   under the conformal 
transformation  $g_{\mu\nu}(x)\to    \Omega^{2}(x)g_{\mu\nu}(x)$ where
$\Omega^{2}(x)$   is   a    finite, non-vanishing,   continuous   real
function. It therefore encompasses the  largest  symmetry group
which keep the  light cones invariant i.e.  the 15 parameter conformal group
(which includes the 10 parameter Poincar\'e group). Many important results have been obtained for conformal
gravity. It has been shown that Birkhoff's theorem is valid for conformal gravity \cite{Riegert2}. The 
linearized equations about 
flat space-time have also been obtained \cite{Riegert1}.
An important result, called the zero-energy theorem, was obtained for conformal gravity \cite{Boulware}.
This theorem states that for the special case of an asymptotically flat space-time 
the total energy is zero. However, we will see in section 2, that far from a localized 
source the metric for conformal gravity is conformal to flat and not flat. Hence, the zero-energy theorem 
only applies to cases where asymptotically flat space-time is imposed as a boundary
condition. Interest in conformal gravity was
rekindled in the 
early 90's after the metric exterior to a static spherically symmetric source
was 
obtained \cite{Mannheimb}. For a metric in the standard form 
\beq
d\tau^{2}=B(r)\,dt^2 - A(r)\,dr^2 -r^{2}\left(d\theta^2 + 
\sin^{2}\theta\,d\varphi^{2}\right) 
\eeq{ds2b}
the static spherically symmetric
vacuum solutions are \cite{Riegert2,Mannheimb} 
\beq B(r)=A^{-1}(r)=1-\dfrac{(2 -
3\gamma \beta)\beta}{r} -3\beta\gamma + \gamma\,r  -kr^2 
\eeq{solutionb}
where $\beta$,$\gamma$ and $k$ are constants. The above solution is only valid
up to a conformal factor. Constraints from phenomenology imply that $\gamma, k$ and 
$\gamma\beta <<1$ (see \cite{Mannheimb,Mann1b,Ederyb}).  The constant
$\gamma\beta$ is usually negligible but for the purposes of the next section we
include it here.  In subsequent sections it will be dropped. 

Mannheim and Kazanas used the above metric solution to fit galactic rotation 
curves without recourse to dark matter i.e. they wanted to verify whether the linear 
$\gamma r$ term could replace dark matter in explaining the rotation curves. They had some success in these
fittings but the deflection of light which was later calculated \cite{Ederyb, Walker} was   
incompatible with the fitting of galactic rotation curves. It was shown \cite{Ederyb} that at large
distances, non-relativistic  massive particles and light behaved in opposite 
ways i.e. if the former was attracted to the source the latter would be repelled and
vice versa. Hence, 
the theory could not simultaneously explain galactic rotation curves and the
observed deflection of light in galaxies. In one important respect, the calculation of the deflection of 
light in conformal gravity is less ambiguous then the calculation of galactic rotation curves. 
The fitting of galactic
rotation 
curves requires one to fix the conformal factor because massive
geodesics are not conformally invariant. The conformal factor is chosen to fit
experiments but there is no theoretical justification for choosing one conformal
factor over another. In contrast, there is no need to specify any confomal
factor for null geodesics because they are conformally invariant. It follows that causal analysis and the 
calculation of the deflection of light can be carried out without specifying any conformal factor. 
It is therefore worthwhile to investigate the 
causal structure of the metric \reff{ds2b} with solution \reff{solutionb}. We find the coordinate 
transformations that render  
the metric in a form which is manifestly
conformal to flat. The causal structure is then analysed 
using Penrose diagrams and we identify which space-times allow a calculation of the scattering of light i.e. 
which space-times allow light to approach the source from infinity. The trajectories and 
deflection of light for these space-times is then calculated. We begin by investigating some of the
geometrical properties of solution \reff{solutionb}. 
  
\section{CURVATURE AND RELATED TENSORS}

Curvature scalars are invariant under coordinate transformations and
therefore are useful for detecting physical singularities. In contrast, the
metric (which is coordinate dependent) may have a coordinate singularity which
is not a physical singularity (the classic example is the coordinate
singularity at the Schwarzschild radius $r=2m$ 
which is not a physical singularity but a horizon). The metric under study is 
\begin{eqnarray}
ds^2 &=& (1 - 3\gamma\beta - \beta(2-3\gamma\beta)/r + \gamma r - k r^{2}) dt^2
\nonumber\\ 
&-& \dfrac{dr^2}{1-3\gamma\beta -\beta(2-3\gamma\beta)/r + \gamma r
-k r^{2}} - r^2 d\Omega^2 .  
\label{metricb}
\end{eqnarray}
The curvature scalar, $R\equiv R_{\mu\nu}g^{\mu\nu}$, for the general metric
\reff{ds2b} with $B(r)=A^{-1}(r)$ can readily be calculated and yields 
$R= B^{''} + 4B'/r +2B/r^2 -2/r^2$ where a prime denotes differentition with
respect to $r$. 
The curvature scalar for the metric \reff{metricb} is equal to
\beq
R= 6\gamma/r -6\gamma\beta/r^2 -12k .
\eeq{riccib}
The first two terms in the curvature scalar are singular at $r=0$ and therefore
the 
space-time described by the metric \reff{metricb} has a physical singularity at
$r=0$. Since the curvature scalar is a linear inhomogeneous function of $B(r)$ it 
follows
that each term that appears in \reff{riccib} can be traced back to a term in $B(r)$.
Therefore the two singular terms are due to the $\gamma r$ and the constant
$3\beta\gamma$ term in the metric respectively.   
The singularity due to the constant $3\beta\gamma$ term is a conical
singularity. This can be shown by considering the metric \reff{metricb} with
only the constant term present i.e. 
\beq
ds^{2}= (1-3\gamma\beta) dt^2 - 1/(1-3\gamma\beta)dr^2 -r^2 (d\theta^2 - \sin^2
\theta d\varphi^2 ). 
\eeq{flat}
This metric exhibits a conical singularity, the ratio of the area of a sphere at
coordinate radius $r$ to the proper radius squared $r^2/(1-3\gamma\beta)$ is the
constant $4\pi (1-3\gamma\beta)\ne 4\pi$.  Correspondingly, the deflection of
light is given by the angular defect in the scattering two plane,
$3\pi\gamma\beta$ in the limit $\gamma\beta <<1$.

The singularity due to the $\gamma r$ term can be analyzed by studying 
the metric 
\beq
d s^2 = (1+\gamma r) dt^2 - 1/(1+\gamma r) dr^2 - r^2 d\Omega^2 
\eeq{gamma}
obtained by setting $\beta=k=0$ in the metric \reff{metricb}. It is not
apparent that the space-time described by the above metric has a
singularity at $r=0$. In fact,
at first glance, it seems that the metric approaches Minkowski space-time as $r$
approaches zero! A singularity at $r=0$ is made apparent by
rewriting the metric \reff{gamma} for small $r$ i.e. $\gamma r <<1$. The metric
\reff{gamma} then takes the form   
\begin{eqnarray}
ds^2 &=& (dt^2 - dr^2 - r^2 d\Omega^2)  + \gamma r(dt^2 + dr^2) \nonumber \\
&=& (dt^2 - \sum_{i=1}^{3} dx_i^{2}) + \gamma \left( \sum_{i=1}^{3} 
\dfrac{ x_{i}^{2} dx_i^{2}}{r} + \sum_{i<j} \dfrac{2 x_{i}x_{j}}{r} dx_{i}
dx_{j} +  
r dt^2 \right)\nonumber\\
&=& (\eta_{\mu\nu} +h_{\mu\nu} )dx^\mu dx^\nu .
\label{metric2b}
\end{eqnarray}
The metric has therefore been decomposed into Minkowski space-time plus an
additional small term. Although neither term is singular, the derivatives of
$h_{\mu\nu}$ are singular at $r=0$.  The connection and the Riemann tensor are
constructed out of these derivatives and the inverse metric $g^{\mu\nu}$ (which
is not singular) subsequently giving rise to physical singularities at $r=0$ in
the space-time.   

Besides the physical singularity at $r=0$, the curvature 
scalar reveals another interesting feature of the space-time. As $r$ tends to 
infinity, the curvature scalar does not vanish but approaches the value $-12k$.
The original  
metric \reff{metricb} therefore describes a space-time where the region 
far from the source i.e. the background,    
is not flat but of constant four-curvature. It will later be shown that this 
constant four-curvature background is actually conformal to flat. 

We now exhibit the Riemann tensor, $R_{\hat\mu\hat\nu\hat\sigma\hat\tau}$, 
in an orthonormal basis for the metric
\reff{metricb}. 
Its non-vanishing components are  
\begin{eqnarray} 
R_{\hat r \hat\theta \hat r \hat\theta}&=& R_{\hat r \hat\phi\hat r \hat\phi} = \dfrac{\beta(2 - 3\gamma\beta)}{2r^3} + 
\dfrac{\gamma}{2r} -k \,, \quad \, R_{\hat r\hat t\hat r\hat t}=\dfrac{\beta(2 - 3\gamma\beta)}{r^3} + 
k \nonumber\\ \,R_{\hat\theta \hat\phi \hat\theta \hat\phi}&=& 
\dfrac{-\beta(2 - 3\gamma\beta)}{r^3} - \dfrac{3\gamma\beta}{r^2}+
\dfrac{\gamma}{r} -k \,, \nonumber \\ R_{\hat\theta\hat t \hat\theta\hat t} &=& 
R_{\hat\phi\hat t \hat\phi\hat t} = 
\dfrac{-\beta(2-3\gamma\beta)}{2r^3}-\dfrac{\gamma}{2r} + k .
\label{riemannb}
\end{eqnarray}
where other non-vanishing components related to the above by symmetry are
not shown. Clearly, the Riemann tensor and the scalars constructed from it, for
example the  Riemann tensor squared, diverge at $r=0$. We see that terms in 
the metric
containing either $\beta$, $\gamma$ or both 
contribute a physical singularity at $r=0$ and represent point-like sources.  The 
$k r^{2}$ term contributes
a constant $\pm k$ to the Riemmann tensor and therefore the components of 
the Riemann tensor do not vanish as $r$ 
tends to infinity i.e.  the space-time is not flat at infinity. The curvature 
scalar and Riemann tensor have revealed that the metric 
\reff{metricb} represents point sources localized at $r=0$ which are embedded in a constant four-curvature 
background . 

Let us now compute the 
Weyl tensor
$C_{\hat\mu\hat\nu\hat\sigma\hat\tau}$ for the metric \reff{metricb}. This tensor is
useful because the requirement that a space-time be conformal to flat
is that the components of the Weyl tensor vanish. The components of
the Weyl tensor in an orthonormal basis are
\begin{eqnarray}
 C_{\hat r \hat\theta \hat r \hat\theta}&=& C_{\hat r \hat\phi \hat r \hat\phi} =
\dfrac{\beta(2-3\gamma\beta)}{2r^3} +  \dfrac{\gamma\beta}{2r^2}\nonumber\\ 
C_{\hat r\hat t\hat r\hat t}&=&\dfrac{\beta(2-3\gamma\beta)}{r^3}+\dfrac{\gamma\beta}{r^2}
\quad C_{\hat\theta\hat\phi\hat \theta \hat\phi}= 
\dfrac{-\beta(2-3\gamma\beta)}{r^3}-\dfrac{\gamma\beta}{r^2} \nonumber\\
C_{\hat\theta\hat t \hat\theta\hat t} &=&
C_{\hat\phi\hat t \hat\phi\hat t} = \dfrac{-\beta(2-3\gamma\beta) }{2r^3} -  \dfrac{\gamma\beta}{2r^2} . 
\label{weylb}
\end{eqnarray} 
where components related to the above by symmetry are not shown. The
Weyl tensor is zero when  
$\beta=0$ or when $r$ approaches infinity. Under these conditions the original 
metric \reff{metricb} reduces to  
\beq 
ds^2=(1+\gamma r -k r^2) dt^2 - 1/(1+\gamma r -k r^2) dr^2 - r^2 d\Omega^2.
\eeq{metric3b} 
The above metric is conformal to flat and it
describes the $\beta=0$ or very large $r$  
limit of the original metric \reff{metricb}.    
We will analyze the conformally flat metric \reff{metric3b} in detail in the
next 
sections, to understand the causal structure of the original metric at very large
radii and in particular to verify whether light has scattering trajectories.   

\section{COORDINATE TRANSFORMATIONS}

In the original $r,t$ coordinates, the components of the metric \reff{metric3b}
change sign at the roots of the polynomial $1+\gamma r -kr^2=0$.
These coordinates are therefore not the most convenient to analyze the causal structure.  
Our task in this section
will 
be to rewrite the conformally flat metric \reff{metric3b} in coordinates where
the 
conformal flatness is manifest i.e. in a form where the metric is a conformal
factor 
times the Minkowski metric.  The effort spent in obtaining the new coordinates
is 
rewarded by having the metric in a form that has the same causal structure as
that of Minkowki space-time i.e. null geodesics do not depend on the conformal
factor and therefore the light cones are drawn at $45^{0}$ to the horizontal
axis 
as in Minkowski space-time. There are constraints on the new coordinates when
one transforms from the old to the new coordinates. Therefore, the causal
structure of the conformally flat space-time is analyzed in the new coordinates
as a  {\it patch} in Minkowski space-time. 


We  now perform the coordinate transformation from the $r,t$ coordinates to a 
new set of coordinates $\rho, \tau$  where the metric \reff{metric3b}
is written in a form which is manifestly conformal to flat.  We write
\begin{eqnarray} 
ds^{2} &=& (1+\gamma r -k r^2) dt^2 - 1/(1+\gamma r -k r^2) dr^2 - r^{2} 
(d\theta^2 +  \sin^{2}\theta\,d\varphi^{2})\nonumber\\ &=
&\Omega^{2}(\rho,\tau)\left[d\tau^2 - d\rho^2 - 
\rho^{2} (d\theta^2 +  \sin^{2}\theta\,d\varphi^{2}) \right]
\label{conformalb}
\end{eqnarray}
where $\tau$ and $\rho$ are the new coordinates and $\Omega(\rho,\tau)$ is the 
conformal factor.  We therefore have the following two relations:
 \beq
r=\rho\,\, \Omega 
\eeq{firstb}
\beq
 \Omega^{2} (d\tau^2 - d\rho^2 ) =  (1+\gamma r -k r^2) dt^2 - 
1/(1+\gamma r -k r^2) dr^2.\
\eeq{omegb}
The coordinates $r$ and $t$ are now functions of both $\rho$ and $\tau$ so that 
$dr=r^{\prime }d\rho + \dot{r} d\tau$ and $dt = 
t^{\prime}  d\rho + \dot{t} d\tau$ where
a prime and dot on $r$ and $t$ represent partial derivatives with respect to
$\rho$ and $\tau$ respectively.  Equations \reff{firstb} and \reff{omegb}  lead
to the 
following three partial differential equations
 \begin{eqnarray}
 (1+\gamma r -k r^2)t^{\prime} \dot{t}   - \dfrac{r^{\prime}  \dot{r}}{1+
\gamma r -k r^2 } &=& 0 \\  
(1+\gamma r -k r^2) \dot{t}^2   - \dfrac{ \dot{r} ^2}{1+\gamma r -k r^2} &= &
\dfrac{ r^{2}}{\rho^2}\\
(1+\gamma r -k r^2){t^{\prime}}^2  - \dfrac{{r^{\prime}}^2}{1+\gamma r -k r^2} 
&=& - \dfrac{ r^{2}}{\rho^2} .
\label{threeb}
\end{eqnarray}
We can eliminate $t$ from the above three equations 
to obtain two partial differential equations for $r$:  
 \beq
\dfrac{r^{2}}{\rho^{2} ( \dot{r} + r^{\prime})} = f(\tau-\rho)
\eeq{ftaub}
\beq
{r^{\prime}}^2 -  \dot{r} ^2 =\dfrac{ r^{2} (1+\gamma r -k r^2)}{\rho^{2}}
\eeq{bubbleb}
where $ f(\tau-\rho)$ is an arbitrary function of $\tau-\rho$. 
To solve the above two equations for $r$, 
it is convenient to introduce two new coordinates $u$ and $v$ 
related to $\tau$ and $\rho$ by
\beq
u= \tau-\rho \quad ;\quad v=\tau+ \rho \,.
\eeq{uvb}
 In $u,v$ coordinates \reff{ftaub} reduces to
\beq
\dfrac{2r^2}{(v-u)^2 \,\partial r /\partial v}=f(u).
\eeq{ftiltb}
The solution to the above equation is 
\beq
r=\dfrac{f(u)(v-u)}{2 + h(u)(v-u)}
\eeq{ftau2b}
where $h(u)$ is an arbitrary function of $u$.
Substituting \reff{ftiltb} into \reff{bubbleb} one obtains
\beq
-2\int \dfrac{dr}{1+\gamma r -k r^2} =\int f(u) du= g(u) + p(v)
\eeq{bubble2b}
where $dg(u)/du = f(u)$ and $p(v)$ is an 
arbitrary function of $v$.  The solution to the above equation depends on 
whether the polynomial
$1+\gamma r -k r^2$ has roots or not.  If the polynomial has roots the integral 
of $1/(1+\gamma r -kr^2)$ is given by
\beq
\dfrac{-1}{k(r_{+}-r_{-})}\ln\left|\dfrac{r-r_{+}}{r-r_{-}}\right|\,; 
\quad  k> -\dfrac{\gamma^2}{4}
\eeq{roots2b}
where the two roots $r_{+}$ and $r_{-}$, which can have negative values, are
given by  
\beq
r_{\pm} = \dfrac{\gamma}{2k} \pm \sqrt{\dfrac{\gamma^2}{4k^2} + \dfrac{1}{k}}. 
\eeq{rootsb}
If the polynomial has no roots the integral of $1/(1+\gamma r -kr^2)$ is given
by  
\beq
\left(-k-\gamma^{2}/4 \right)^{-1/2} \arctan\left(
\dfrac{-kr +\gamma/2}{\sqrt{-k-\gamma^{2}/4}}\right)\,; \quad k< -\gamma^{2}/4.
\eeq{norootsb}
We now solve \reff{bubble2b} separately for each of the two cases i.e. case
1: 
polynomial has roots and case 2: polynomial has no roots.   

\vspace{1em}\vspace{1em}\noindent{\it Case 1: Roots at $r_{\pm}$} 
\vspace{1em}

Substituting \reff{roots2b} for the integral in \reff{bubble2b} one obtains
\beq
r= \dfrac{r_{+} \pm r_{-}e^{k(r_{+}-r_{-})( g(u) + p(v))/2}}{1 \pm
e^{k(r_{+}-r_{-}) 
( g(u) + p(v))/2}} 
\eeq{rot3b}
where the negative sign corresponds to the region where 
$\infty>r>r_{+}$ and $0<r<r_{-}$ whereas the plus sign corresponds to the
region  b
where $r_{+} >r >r_{-}$. 
We now equate $r$ in \reff{ftau2b} to $r$ in \reff{rot3b}. Note that
$f(u)$ in \reff{ftau2b} is $g^{\prime}(u)\equiv dg(u)/du$.  One obtains the 
following equality
\beq
\dfrac{g^{\prime}(u)e^{-k(r_{+}-r_{-})( g(u) + p(v))/2}}
{ r_{+}e^{-k(r_{+} -r_{-})( g(u) + p(v))/2} \pm r_{-} } \pm 
\dfrac{g^{\prime}(u)e^{k(r_{+}-r_{-})( g(u) + p(v))/2}}{  r_{+ }  
\pm r_{-} e^{k(r_{+}-r_{-})( g(u) + p(v))/2}}= \dfrac{2}{v-u} + h(u).
\eeq{longb}   
After integrating the above equation and performing some algebraic
manipulations we obtain   
\beq
\ln\left(r_{+}e^{-k(r_{+}-r_{-})( g(u) + p(v))/4} \pm r_{-} e^{k(r_{+}-r_{-})(
g(u) + p(v))/4}\right)=\ln(v-u) + S(u) +  T(v) 
\eeq{longerb}
where $S(u)$(related to $h(u)$) and $T(v)$ are arbitrary functions of $u$ and
$v$ repectively. After exponentiating both sides 
\reff{longerb} reduces to  
\beq
r_{+}e^{-k(r_{+}-r_{-})P(v)/2} \pm  r_{-} e^{k(r_{+}-r_{-})g(u)/2} = (v-u)
A(u)B(v). 
\eeq{againb}
The functions $A(u),B(v),g(u)$ and $p(v)$ are arbitrary functions of $u$ and
$v$  
and we can therefore write the above equation as
\beq
(v-u)A(u)B(v) = N(v) + M(u)
\eeq{finalb}
where all the functions above are arbitrary functions of $u$ and $v$.  The 
coordinate $r$, given by \reff{rot3b}, can be expressed in terms of the
functions $M(u)$ and $N(v)$ i.e. 
\beq
r=\dfrac{ r_{+}\, r_{-}\left(M(u) + N(v)\right)}{ r_{-}\,N(v) + r_{+}\, M(u)} 
\eeq{rfinalb}
where the above is valid for the entire region $\infty>r>0$.  Fortunately,
equation  
\reff{finalb} can be solved algebraically.  We 
arrive also at \reff{finalb} in case 2 and therefore postpone finding its
solution until case 2 is completed.

\vspace{1em}\noindent{\it Case 2: Polynomial has no roots} 
\vspace{1em}

We proceed in a fashion similar to case 1. Substituting \reff{norootsb} for the 
integral in \reff{bubble2b} one obtains
\beq
r=\dfrac{1}{kc}\tan\left[(g(u)+p(v))/2c \right] +\dfrac{\gamma}{2k}
\eeq{kcb}
where $c\equiv \,-1/\sqrt{-k-\gamma^2 /4}$.  We now equate $r$ in 
\reff{ftau2b} to $r$ in \reff{kcb}.  After integration one obtains
\beq
\ln\left[\cos\left((g(u)+p(v))/2c\right)\,\gamma\,c/2
+\sin\left((g(u)+p(v))/2c\right) 
\right] = \ln(v-u) + S(u) +  T(v)
\eeq{sincosb}
where $S(u)$ and $T(v)$ are arbitrary functions.  After algebraic manipulations 
one obtains
\beq
\tan(g(u)/2c) +\dfrac{\sin(p(v)/2c) +\cos(p(v)/2c)\,\gamma\,c/2}{\sin(p(v)/2c)\,
\gamma\,c/2 +\cos(p(v)/2c)} = (v-u)A(u)B(v).
\eeq{serfb}
We therefore obtain the same equation as in case 1 i.e.
\[
(v-u)A(u)B(v) = N(v) + M(u).
\]  
In terms of the functions $M(u)$ and $N(v)$, the coordinate $r$, given by 
\reff{kcb} is
\beq
r=\dfrac{-c \,(M(u)+N(v))}{1+ (M(u)+N(v))\gamma\,c/2 -M(u)N(v)}.
\eeq{r2b}
Though we arrive at the same equation \reff{finalb}, the coordinate $r$ in case
1 and case 2 are obviously not the same.  

We now solve \reff{finalb} and discuss its physical significance. The right
hand side of the equation does not contain any mixed terms of $u$ and $v$ and 
therefore the
mixed terms on the left hand side must vanish. We write $A(u)$ as
\beq
A(u) =A_{0} + a(u^{\prime})
\eeq{Ab}
where $A_{0}= A(u_{0})$ is a constant and $ a(u^{\prime})$ is a function of
$u^{\prime}\equiv u-u_{0}$ which vanishes at $u^{\prime} =0$. Similarly
\beq
B(v) = B_{0} + b(v^{\prime}).
\eeq{Bb}
With $A(u)$ and $B(v)$ given above, the left hand side of \reff{finalb}
yields 
\beq
\left(v^{\prime} -u^{\prime} + C_{0}\right) \left(A_{0} B_{0} +  A_{0}
b(v^{\prime}) + 
  B_{0} a(u^{\prime}) + a(u^{\prime}) b(v^{\prime})\right)
\eeq{vub}
where $C_{0} = v_{0}-u_{0}$ is a constant. The mixed terms must vanish and we 
obtain the following equation
\beq
v^{\prime} a(u^{\prime})B_{0} + v^{\prime} a(u^{\prime})  b(v^{\prime})  
-u^{\prime}b(v^{\prime}) A_{0}  -u^{\prime} a(u^{\prime})b(v^{\prime}) + 
C_{0} a(u^{\prime})b(v^{\prime})= 0.
\eeq{mixedb}
The solution to \reff{mixedb} is obtained by separating the variables i.e. 
\beq
b(v^{\prime})= \dfrac{-v^{\prime}B_{0}}{C_{0} + v^{\prime} - u^{\prime}\left(1+
A_{0}/ a(u^{\prime})\right)}.
\eeq{secb}
The function $b(v^{\prime})$ is a function of $v^{\prime}$ only and therefore
the term $ u^{\prime}\left(1+A_{0}/ a(u^{\prime})\right) $ must be a constant 
(call it $D$). We therefore obtain the following solutions
\beq
a(u^{\prime})= \dfrac{-u^{\prime}A_{0}}{D + u^{\prime}} \,;\quad b(v^{\prime})= 
\dfrac{-v^{\prime}B_{0}}{C_{0} + D+ v^{\prime}}.
\eeq{solt2b}
The solution to 
$ ( v - u )\,A(u)\,B(v) = M(u) + N(v)$ is  therefore   
\beq
A(u) = \dfrac{A}{B+ u},\,B(v) = \dfrac{C}{B+v},\,M(u)=\dfrac{-AC\,u}{B(B+
u)},\,N(v) 
= \dfrac{AC\,v}{B(B+v)}
\eeq{solu2b}
where the solution \reff{solt2b} was substituted into equations \reff{Ab} and 
\reff{Bb} and the
quantities $A,B$ and $C$ are constants related to the constants 
$A_{0},B_{0}, C_{0}$ and $D$.  
With the above solution we can finally obtain the coordinate $r$. For case 1, 
$r$ is given by \reff{rfinalb} and yields
\beq
r=\dfrac{r_{+}r_{-}(v-u)}{vr_{-}(1+u/B)-ur_{+}(1+v/B)} . 
\eeq{general}
For case 2, $r$ is given by \reff{r2b} and yields 
\beq
r= \dfrac{-cAC (v-u)}{(B+u)(B+v) - \gamma cAC(v-u)/2 + (AC/B)^{2}uv} .
\eeq{general2}
It is worth mentioning that \reff{finalb} is invariant under the the
following  
coordinate transformation  
\beq
u\to \dfrac{Au}{B+u} \quad  v\to \dfrac{Av}{B+v}  
\eeq{specialb} 
where $A$ and $B$ are arbitrary constants. This is because these transformations form a 
subgroup of the special conformal transformations i.e. that leave the equation
defining null surfaces 
\beq
ds^2 = d\tau^2 - d\rho^2 -\rho^2 d\Omega^2 = dudv - 
\left(\dfrac{v-u}{2}\right)^2 d\Omega^2 =0
\eeq{invb}
invariant (where $u$ and $v$ are related to $\tau$ and $\rho$ via
\reff{uvb}).  
The transformations form only a subgroup of the 
15 parameter conformal group because two coordinates are not involved in the 
transformation. We see that equations \reff{finalb} 
and \reff{invb} share the same symmetries. There are two more transformations 
that leave equations \reff{finalb} and \reff{invb} invariant. These are 
\beq
\word{space inversion:}\,u \to v \,,\quad v \to u \quad i.e.\quad \rho \to -\rho
\eeq{spaceb}
\beq
\,\,\,\,\word{time reversal:}\, u\to -v \,, \quad v \to -u  \quad i.e.\quad
\tau \to -\tau. 
\eeq{timeb}    
Of course, these can be combined with transformations \reff{specialb}.


\section{PENROSE DIAGRAMS}

The causal structure of the  conformally flat metric \reff{metric3b} will
now be  
analyzed for different choices of $\gamma$ and $k$ 
in the ``conformally flat coordinates" $u$ and $v$ (or $\tau $ and
$\rho$). 
The possible choices of $\gamma$ and $k$ are the following
\beq
 \begin{array}{l} a) \,\,k>0 : \,  r_{+} >0,\, r_{-}<0 \\
b) \,\, -\gamma^2 /4<k<0 \word{and} \gamma<0 : \, r_{+} >0,\, r_{-}>0 \\ 
c)\,\, -\gamma^2 /4<k<0 \word{and}\gamma>0 :\,r_{+} <0,\, r_{-}<0\\
d)\,\,k<-\gamma^2 /4 : \word{no roots}
\end{array}
\eeq{choiceb}
\begin{figure}
\vspace{-2.cm}
\leavevmode
\epsfxsize=0pt\epsfbox{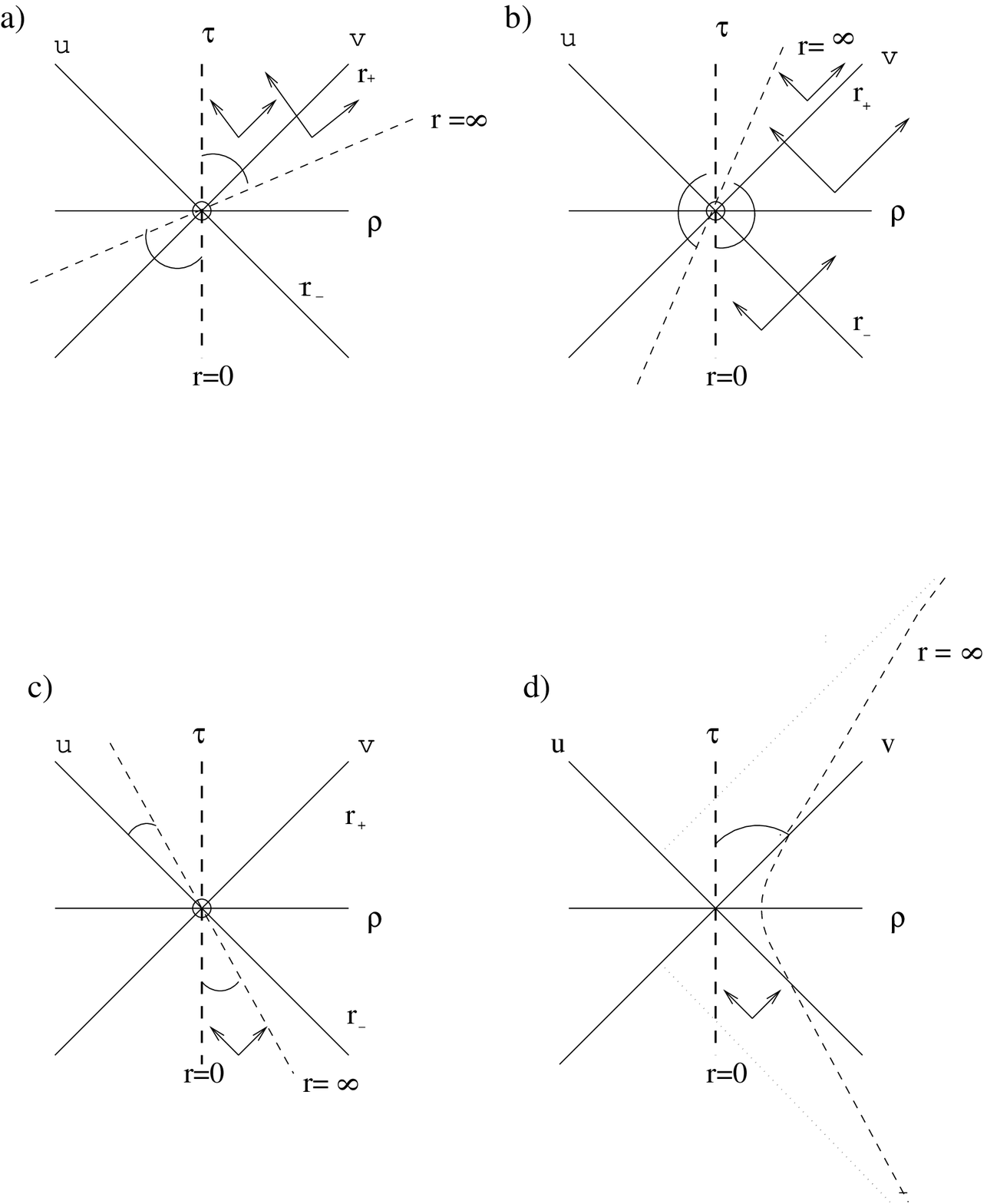}
\vspace{-2.cm}
\caption{\label{Penrose} Penrose diagrams for four different space-times: 
a) horizon at $r_{+}$  b) horizon at $r_{-}$ and $r_{+}$  c) roots $r_{\pm}$
are  
negative, no horizon
d) no roots, no horizon} 
\end{figure}

Altogether there are four cases to consider and a Penrose diagram has been
drawn for each showing the axes of both the $u,v$ and $\rho,\tau$ coordinates
(see figure \ref{Penrose}). The causal analysis proceeds as in Minkowski
space-time except that only a patch of the $u,v$ (or $\rho,\tau$ ) coordinates
are 
allowed. This is due to the condition that in the original $r,t$ coordinates the
radius $r$ must be positive.  In every diagram the singularity at $r=0$ is
shown in 
bold as a vertical dashed line occuring at  $u=v$ or $\rho =0$.  We draw a
circle 
at the point $u=v=0$ in the first three diagrams to show that $r$ is
indeterminate 
at that point i.e. the origin does not correspond to any one specific value of
$r$ 
but depends on the limit with which one approaches it.  If a line crosses the
origin, the value of $r$ at the origin will depend on the slope of that line. 
For all 
the four diagrams, the line at $r=\infty$ is represented by a dashed line. The
region where $r$ is positive and runs from the singularity at $r=0$ to the
dashed 
line at $r=\infty$ is shown by an arc ( there is a second arc that is shown that
represents an identical patch but with time running the opposite direction). 
The 
lines with arrows represent radial null geodesics i.e. the light cones. 
All the diagrams are drawn for the special case where the constant $B$
in \reff{general} and \reff{general2} approaches infinity.
Diagrams a),
b) 
and c) represent the case with roots at $r_{\pm}$  and therefore the coordinate
$r$ is given by \reff{general}.   One
obtains 
the following features for all three diagrams: $r_{+}$ is a $45^{0}$ line at
$u=0$ 
and $r_{-}$ is a $-45^{0}$ line  at $v=0$, lines of constant $r$ are simply
straight 
lines that go through the origin and the radius $r$ approaches $r_{+}$ as $v\to
\infty$ (this is not shown on the diagrams to avoid clutter).  Diagram d)
represents the case with no roots and therefore the coordinate $r$ is given by
\reff{general2}. 
Lines of constant $r$ are hyperbolas and do not go
through 
the origin.  Unlike the first three diagrams, $r$ is zero at the origin and
therefore a 
physical singularity exists at the origin.  We therefore do not draw a circle
at the 
origin as with the other three diagrams. 

We now investigate the causal structure of
all four diagrams. In diagram a), the case of $k>0$, $r_{+}$ is a horizon
because 
light between $r=0$ and $r=r_{+}$ either ends at the singularity or at $r_{+}$ i.e. the  $r_{+}$ at 
$v\to\infty$ which is not shown on the diagram. Note that there exists no point from which
light 
can reach infinity. Clearly, there are no scattering states for the
space-time described by diagram a).  In diagram b), both $r_{-}$ and $r_{+}$ are positive and act
as  horizons. Light at a radius greater than $r_{-}$ cannot cross the $r_{-}$
line. Light  between
$r_{+}$ and 
infinity cannot cross the $r_{+}$ line and is trapped between these two values.  No relevant scattering can therefore take place i.e. the radius of closest
approach is greater than $r_{+}$ (which is a radius on cosmological scales since $\gamma /k$ is of that magnitude by definition).   
In diagram c), there are no
horizons i.e. both $r_{+}$ and $r_{-}$ are in the negative $r$ region and are
outside the patch shown by the arc.  In this space-time, light at infinity can reach
any radius $r_{0}$ and return back to infinity (see fig.\ref{straight2}b).  Hence, scattering takes place in 
diagram c). In diagram d),
the 
case with no roots, there are no horizons and again light at infinity can reach any
radius and return back to infinity. Scattering therefore occurs in diagram d). Therefore, of the four
possible 
space-times, only those described by diagram c) and d) have scattering.

\section{TRAJECTORIES AND DEFLECTION OF LIGHT}      

We saw in the previous section that light has scattering states only for the
space-times described by diagram c) and d) i.e.  when $0>k>-\gamma^2/4$ with
$\gamma>0$ or $k<-\gamma^{2}/4$ respectively (note that no scattering states
exist for a positive value of $k$).  We can therefore calculate the deflection
of 
light for the two cases above. The deflection of light has already been
calculated 
for the original metric \reff{metricb}. The result obtained is \cite{Ederyb} 
\beq
\dfrac{4\beta}{r_{0}}-\gamma r_{0}
\eeq{defb}
where $r_{0}$ is the point of closest approach. The calculation was done with
the 
approximation that both terms in \reff{defb} are much smaller than one.  At
large $r_{0}$, however, the $\gamma r_{0}$ term is not small and the
approximation is therefore no longer valid.  It is therefore worthwhile to
perform a 
separate calculation for the deflection of light at large $r_{0}$. If $r_{0}$
is large 
enough, we can neglect the $\beta$ term and light will therefore be scattering
in 
a conformally flat space-time.  Before calculating the deflection of light, it
is 
worthwhile to understand how a deflection is possible in a conformally flat
space-time. 

\begin{figure}[ht]
\centering
\mbox{\epsfig{figure= 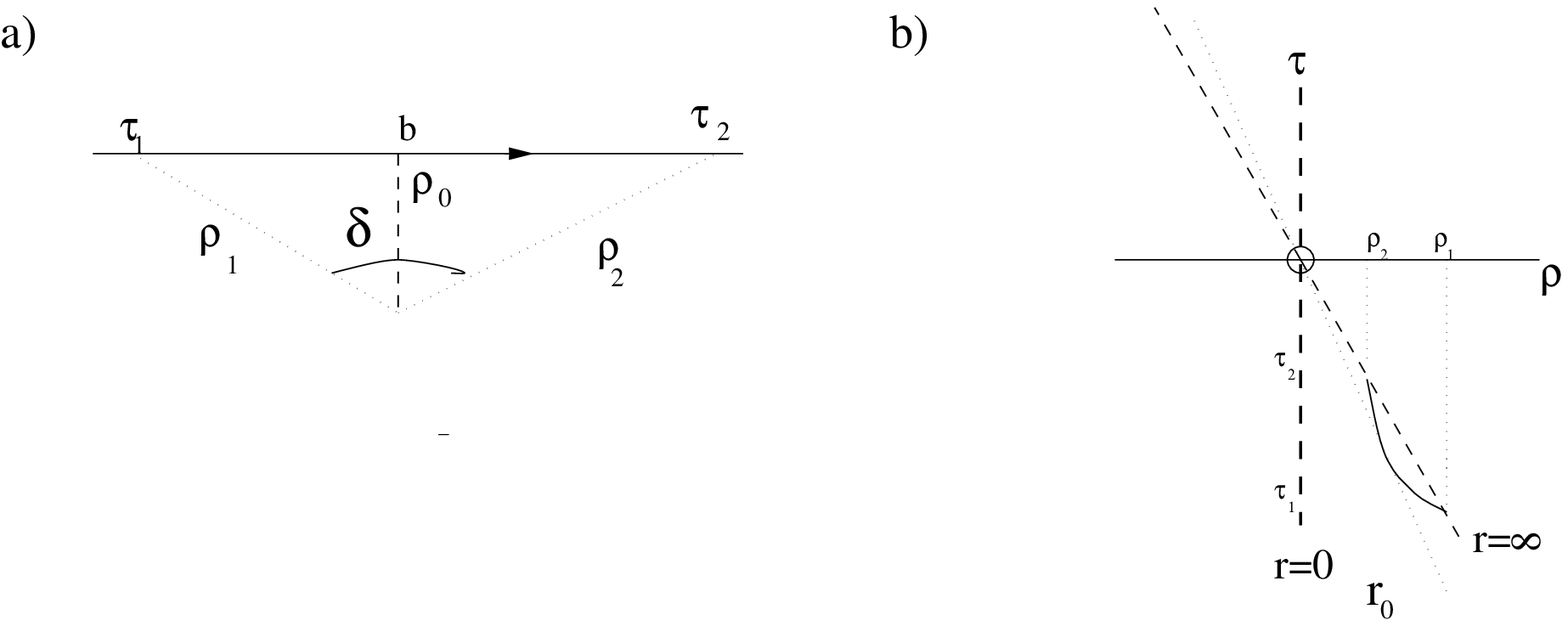,width=15.5cm}}
\vspace{-12.0cm}
\caption{\label{straight2} 
a) Straight line motion from $\rho_{1}, \tau_{1}$ to $\rho_{2},\tau_{2}$. The
angle  
$\delta$ is less than $\pi$.  b) path of light, shown as the curved solid line,
 as it moves from $r=\infty$ at $\rho_{1}$, to $r_{0}$, and  back to $r=\infty$
at  
$\rho_{2}$.} 
\end{figure}

In the ``conformally flat" coordinates $\rho$ and $\tau$, light
moves 
in a straight line as in Minkowski space-time.  How can one then have
scattering? In a scattering process light starts far away from the source,
approaches the source, and then ends up far away from the source.  ``Far away"
means that the sources no longer have any influence on the trajectory of the
light. By looking at the curvature scalar and Riemann tensor we know that the
sources no longer have influence as $r$ approaches infinity. However, if the
coordinate $r$ approaches infinity this does not imply that the coordinate
$\rho$ 
approaches infinity. As can be seen in all 4 diagrams in figure \ref{Penrose},
there 
are finite values of $\rho$ and $\tau$ that correspond to $r=\infty$.  Hence,
when 
light moves from $r=\infty$ to $r=\infty$,  it moves from one finite value of
$\rho$ 
and $\tau$, say $\rho_{1}, \tau_{1}$ to another finite value of $\rho$ and
$\tau$, 
say $\rho_{2}, \tau_{2}$.  Hence, in $\rho$, $\tau$ coordinates light moves in a
straight line but it does not cover the entire line i.e. it covers an angle
$\delta$ less than 
$\pi$ (see figure \ref{straight2}a ). The deflection angle is therefore equal to
$\delta - \pi$.  A straight line in polar coordinates $\rho$ and $\tau$ is
described by the equation
\beq 
\tau=\pm\sqrt{\rho^{2}-\rho_{0}^{2}} + b
\eeq{straight}
where $\rho_{0}$ and $b$ are constants. For the space-time described by the
Penrose diagram figure \ref{Penrose}c, we use the straight line equation
\reff{straight} to draw the path of light as it moves from $\rho_{1}, \tau_{1}$
($r=\infty$), to $r_{0}$ and then to $\rho_{2}, \tau_{2}$ ($r=\infty$).  We have
therefore seen how light can deflect in coordinates where the space-time is
manifestly conformal to flat.

We now calculate the deflection of light. The angle $\varphi$ as a function of
$r$  
for the metric \reff{metric3b} is given by 
(see \cite{Ederyb})
\beq
\varphi(r) = \int\left[1 + \dfrac{\gamma r_{0}}{1+\sin\theta}\right]^{-1/2}
d\theta 
\eeq{deflectb}
where $\sin\theta=r_{0}/r$. We therefore obtain the condition that 
\beq
1 + \dfrac{\gamma r_{0}}{1+\sin\theta}\ge 0.
\eeq{condb}
The above condition is automatically satisfied if $\gamma$ is positive and 
this implies that for positive $\gamma$ light will reach infinity($\theta=0$)
for any  
value of $r_{0}$. 
If $\gamma$ is negative, then condition \reff{condb} implies that 
\beq
r_{0}\le\dfrac{1+\sin\theta}{|\gamma|}\quad ;\quad \gamma \word{negative}. 
\eeq{gammab} 
With the above condition, light can reach infinity($\theta=0$) only
if $r_{0}\le 1/|\gamma|$. If $r_{0}$ is in the range $2/|\gamma|\ge r_{0}>
1/|\gamma|$ then light moves in a closed orbit i.e. a bound state.  Let us
now 
calculate the integral \reff{deflectb}. This yields 
\beq
\varphi(r)=\arcsin\left(\dfrac{r_{0}/r + \gamma r_{0}/2}{1 + \gamma
r_{0}/2}\right)   .
\eeq{def2b}
The deflection from infinity to $r_{0}$ and back to  infinity is
\begin{eqnarray}
\Delta\varphi &=&2\left(\varphi(r_{0})- \varphi(\infty)\right) -\pi \nonumber\\
&=&-2 \arcsin\left(\dfrac{\gamma r_{0}}{2+\gamma r_{0}}\right).
\label{def5b}
\end{eqnarray}
For small deflections (i.e. $\gamma r_{0}<< 1$ ) equation \reff{def5b} reduces to  $-\gamma r_{0}$ in
agreement with \reff{defb}. The deflection is repulsive for a
positive $\gamma$ and attractive for a negative $\gamma$. For positive $\gamma$ the deflection ranges 
from $0$ at $r_{0}=0$ to $-\pi$ at $r_{0}= \infty$ and for a negative $\gamma$ it ranges
from 
$0$ at $r_{0}=0$ to $\pi$ at $r_{0}=1/|\gamma|$ (there are no scattering states for negative $\gamma$ 
when $r_{0}> 1/|\gamma|$). 
Let us now obtain the shape of
the orbits. One can obtain $r$ as a function of $\varphi$ from \reff{def2b}.
This 
yields
\beq
r=\dfrac{-2/\gamma}{1-\left(\dfrac{2 + \gamma r_{0}}{\gamma r_{0}}\right)
\sin\varphi}.
\eeq{orbitb}
This is of course the equation for a conic section in polar coordinates with
eccentricity 
\beq
e= \left|\dfrac{2 + \gamma r_{0}}{\gamma r_{0}}\right| . 
\eeq{eccb}
The shapes are determined by the value of $e$. The orbits we obtain are
\beq
\word{positive}\gamma : \word{hyperbola}(e>1)
\eeq{positb}
\beq
\word{negative}\gamma :\left\{ \begin{array}{l}r_{0}< \dfrac{1}{|\gamma|},
\word{hyperbola}(e>1)\\r_{0}= \dfrac{1}{|\gamma|},\word{parabola}(e=1)\\ 
\dfrac{2}{|\gamma|}>r_{0}> \dfrac{1}{|\gamma|},\word{ellipse}(0<e<1)\\r_{0}= 
\dfrac{2}{|\gamma|},\word{circle}(e=0)
\end{array}\right.\,.
\eeq{shapesb} 

For a positive $\gamma$, the shapes of all orbits are hyperbolas and these
describe scattering states.  For a negative $\gamma$, the shapes of the orbits
depend on the value of $r_{0}$ and bound states as well as  scattering states
can exist.  The ellipse has a minimum value of $r$ which is $r_{min}=r_{0}$ and has a maximum value of 
$r_{max}=r_{0}/(|\gamma|r_{0}-1)$. 
The semi-latus rectum $L$, defined to be the point which occurs at an angle of $\pm
\pi/2$ away from $r_{min}$ is equal to $2/|\gamma|$. All ellipses therefore have the same value 
for $L$ i.e. independent of $r_{0}$. For the negative $\gamma$ case,  
what is observable lies in the region where the trajectory is a hyperbola. 
Values for $\gamma $ have previously been obtained and its inverse is typically of the order of the Hubble length
\cite{Mannheimb, Ederyb}.  The bound states therefore occur on cosmological scales where they can never be
traced to any one particular source.    

\section*{ACKNOWLEDGEMENTS}

We thank Robert Mann for useful discussions and NSERC of Canada and
FCAR du Qu\'ebec for financial support.

\end{document}